\documentclass[a4paper,fleqn,usenatbib]{mnras}

\usepackage{newtxtext,newtxmath}
\usepackage{longtable}

\usepackage[T1]{fontenc}
\usepackage{ae,aecompl}

\usepackage{graphicx}	
\usepackage{amsmath}	
\usepackage{amssymb}	
\usepackage{ulem}
\usepackage{float}
\usepackage{placeins}

\title[VVV-WIT-04]{VVV-WIT-04: an extragalactic variable source caught
  by the VVV Survey\thanks{Based on  observations taken within the ESO
    Public  Surveys  VVV  and   VVVX,  Programme  IDs  179.B-2002  and
    198.B-2004, respectively.}}

\author[R.~K.~Saito et al.]{R.~K.~Saito$^{1}$\thanks{E-mail: saito@astro.ufsc.br},
D.~Minniti$^{2,3,4}$,
V.~D.~Ivanov$^{5}$,
N.~Masetti$^{6,2}$,
M.~G.~Navarro$^{2,7,3}$,
\newauthor
R.~Cid Fernandes$^{1}$,
D.~Ruschel-Dutra$^{1}$,
L.~C.~Smith$^{8, 9}$,
P.~W.~Lucas$^{9}$,
\newauthor
C.~Gonzalez-Fernandez$^{10}$,
R.~Contreras Ramos$^{11, 3}$
\\
\\
$^{1}$Departamento de  F\'{i}sica, Universidade  Federal de
  Santa Catarina, Trindade 88040-900, Florian\'opolis, SC, Brazil\\
$^{2}$Departamento  de  Fisica,  Facultad  de  Ciencias  Exactas,
  Universidad  Andres Bello,  Av.  Fernandez  Concha 700,  Las Condes,\\
  Santiago, Chile\\
$^{3}$Instituto Milenio de Astrof\'isica, Santiago, Chile\\
$^{4}$Vatican Observatory, V00120 Vatican City State, Italy\\
$^{5}$European   Southern   Observatory,  Karl-Schwarszchild-Str.   2,
  D-85748 Garching bei Muenchen, Germany\\
$^{6}$INAF-Osservatorio di Astrofisica e Scienza dello Spazio, Via
  Piero Gobetti 93/3, I-40129 Bologna, Italy\\
$^{7}$ Dipartimento di  Fisica, Universit\`a degli Studi  di Roma ``La
Sapienza'', P.le Aldo Moro, 2, I00185 Rome, Italy\\
$^{8}$  Institute of  Astronomy,  University  of Cambridge,  Madingley
Road, Cambridge, CB3 0HA, UK\\
$^{9}$ Centre for Astrophysics  Research, School of Physics, Astronomy
and Mathematics,  University of Hertfordshire, College  Lane, \\ Hatfield
AL10 9AB, UK\\
$^{10}$  Institute of  Astronomy, University  of Cambridge,  Madingley
Road, Cambridge CB3 0HA, UK 0000-0003-2612-0118\\
$^{11}$  Instituto de  Astrofisica, Pontificia Universidad Catolica  de Chile,
Vicuna Mackenna 4860, Macul, Santiago, Chile
}

\date{Accepted XXX. Received YYY; in original form ZZZ}

\pubyear{2015}

\begin{document}
\label{firstpage}
\pagerange{\pageref{firstpage}--\pageref{lastpage}}
\maketitle

\begin{abstract}

We report the discovery of VVV-WIT-04, a near-infrared variable source
towards  the Galactic  disk located  $\sim$0.2 arcsec  apart from  the
position of  the radio  source PMN J1515-5559.   The object  was found
serendipitously in  the near-IR  data of the  ESO public  survey VISTA
Variables  in the  V\'ia L\'actea  (VVV).   Our analysis  is based  on
variability,  multicolor, and  proper  motion data  from  VVV and  VVV
eXtended   surveys,  complemented   with   archive   data  at   longer
wavelengths. We suggest that VVV-WIT-04 has an extragalactic origin as
the  near-IR  counterpart  of  PMN J1515-5559.  The  $K_{\rm  s}$-band
light-curve of VVV-WIT-04 is highly  variable and consistent with that
of an Optically Violent Variable (OVV) quasar.  The variability in the
near-IR  can be  interpreted  as the  redshifted optical  variability.
Residuals to  the proper motion  varies with the  magnitude suggesting
contamination by a blended source.  Alternative scenarios, including a
transient  event  such as  a  nova  or  supernova,  or even  a  binary
microlensing event are not in agreement with the available data.

\end{abstract}




\section{Introduction}
\label{sec:intro}
 
In the  past years, the  VISTA Variables  in the V\'ia  L\'actea (VVV)
survey  has scanned  our Milky  Way (MW)  galaxy in  the near-infrared
(near-IR)         searching        for         variable        sources
\citep{2010NewA...15..433M,2012A&A...537A.107S}.   VVV is  an European
Southern Observatory  (ESO) variability  survey, focused  on unveiling
the 3-dimensional structure of the Milky Way using distance indicators
such as pulsating  RR Lyrae and Cepheids, as well  as red clump stars.
In     2016     the      complementary     VVV     eXtended     Survey
\citep[VVVX,][]{2018ASSP...51...63M}  started  observations,  widening
the survey  area.  It is  also revisiting the original  VVV footprint,
thus extending  the original  time baseline as  a result  of combining
both the VVV and VVVX datasets.

\begin{figure*}
\centering
\includegraphics[scale=1.15]{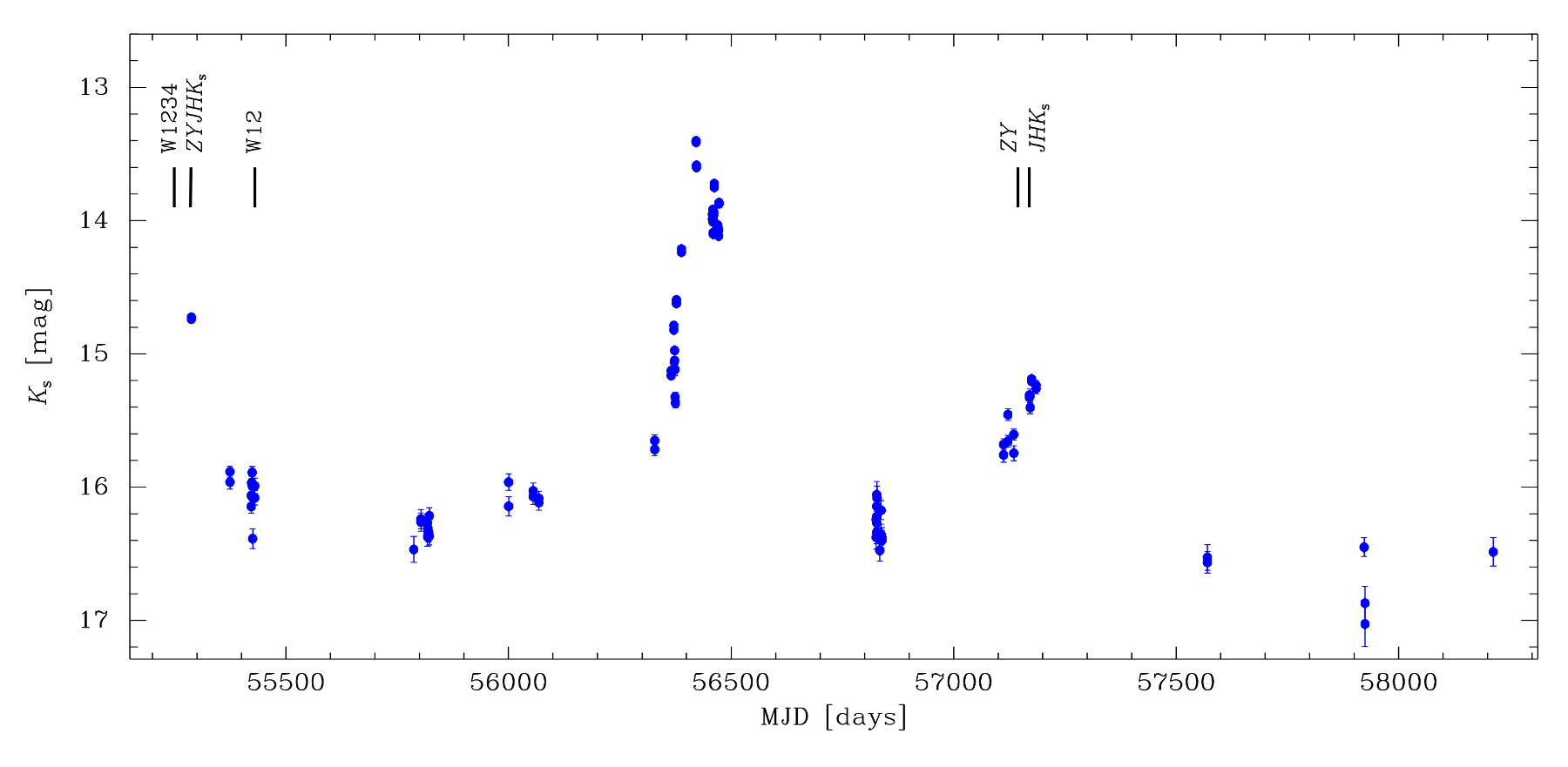}
\caption{$K_{\rm  s}$-band light-curve  of  VVV-WIT-04 combining  data
  from the VVV and VVVX surveys.   There is a total of 111 data-points
  spanning from  March 31, 2010  to April 3,  2018. The two  epochs of
  WISE observations as well as the epochs for the multicolour VVV data
  are marked.  ``W1234''  means WISE observations in  the four filters
  ($W1$,  $W2$,   $W3$  and   $W4$)  while  ``W12''   represents  WISE
  observations only in $W1$ and $W2$.}
\label{fig:lcurve}
\end{figure*}

Besides its main  goal, VVV has also contributed to  the discovery and
study of  variable sources such  as eclipsing binaries,  young stellar
objects,  planetary   transits,  RR   Lyrae  and   Cepheid  variables,
etc. Specifically, a search for transient sources such as microlensing
events and  novae outbursts has resulted  in the discovery of  a large
number    of    new   events    in    the    inner   MW    \cite[e.g.,
][]{2013A&A...554A.123S,2016ATel.8602....1S,2017ApJ...851L..13N,
  2018ApJ...865L...5N}.   Among the  targets  found as  high-amplitude
transient sources in  the VVV data, some caught  our attention because
their behaviour  does not seem to  fit the currently known  classes of
stellar variability.  We named these targets as ``What Is This'' (WIT)
objects.  These rare sources include SN candidates in the MW or behind
it            \citep[VVV-WIT-01             and            VVV-WIT-06,
][]{2012ATel.4041....1M,2017ApJ...849L..23M}  and  a  possible  second
example     of     the     ``Tabby's     star''     \citep[VVV-WIT-07,
][]{2019MNRAS.482.5000S}.

VVV-WIT-04 is a  transient source located $\sim$0.2  arcsec apart from
the position of the radio source PMN J1515$-$5559 in the inner MW disk
\citep{1994ApJS...91..111W},   discovered  in   a  search   for  large
amplitude objects  in the  VVV data  \citep{2015ATel.8456....1S}.  VVV
observations during years 2010$-$2013  showed VVV-WIT-04 increasing in
brightness by  $\Delta K_{\rm s}>2.5$~mag.   Based on the  VVV $K_{\rm
  s}$-band    light-curve     limited    to    the     2013    season,
\cite{2015ATel.8456....1S} suggested  that it  was a Galactic  nova or
even a supernova in a galaxy behind the Milky Way.

Here  we present  an  analysis  of VVV-WIT-04  based  on the  VVV/VVVX
variability, multicolor, and proper  motion data covering 2010$-$2018.
Complementary archive data in the  mid/far infrared and radio aided in
the analysis  and interpretation.  We  suggest that VVV-WIT-04  is the
near-IR  counterpart  of  the   radio-source  PMN  J1515$-$5559.   The
variability in  the near-IR  is consistent  with an  Optically Violent
Variable  (OVV), and  can be  interpreted by  the optical  variability
shifted  towards longer  wavelengths. Alternative  scenarios are  also
discussed, none of which is fully consistent with the available data.

\begin{figure}
\centering \includegraphics[scale=1.02]{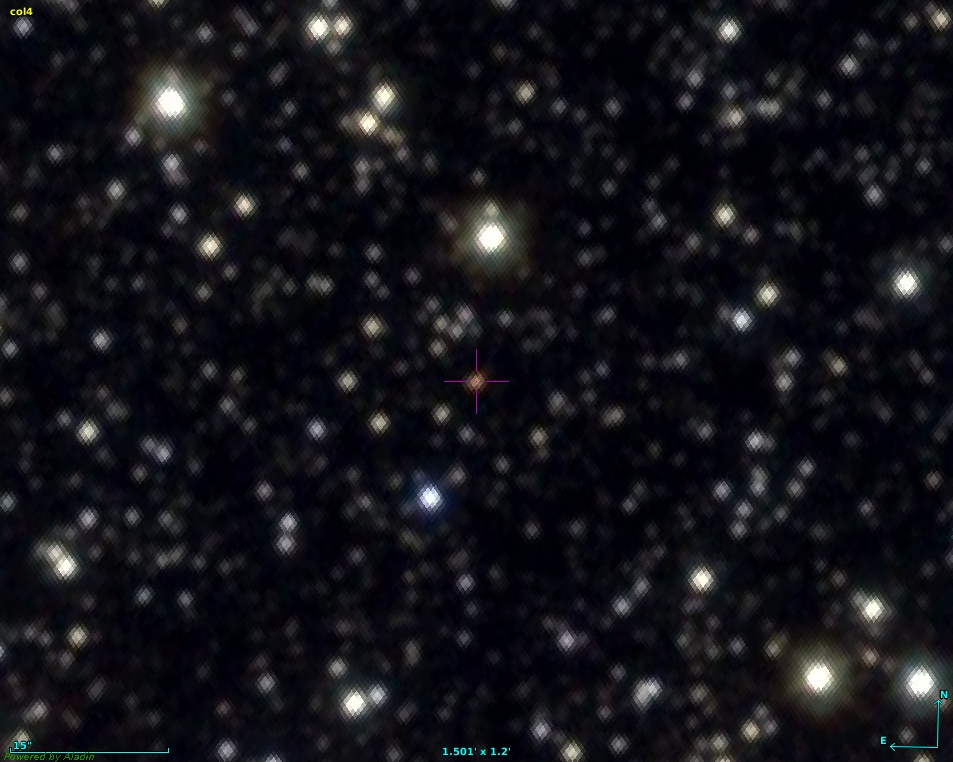}
\caption{VVV $JHK_{\rm s}$ false-color image of VVV-WIT-04 area based
  on observations taken in year 2010 (see Table 1).  The field size is
  $1.5' \times 1,2'$ and oriented  in equatorial coordinates. North is
  towards  the top  and East  towards the  left.  The  reticle at  the
  centre marks  VVV-WIT-04. We  note that the  object is  the reddest
  source in the field.}
\label{fig:image}
\end{figure}

\begin{table*}
\caption[]{Archive data  for VVV-WIT-04.  Observations are  limited to
  long  wavelengths.   The  VVV  $K_{\rm  s}$  epochs  presented  here
  correspond to the ones observed  simultaneously with the $J$ and $H$
  bands.   WISE epochs  and magnitudes  are mean  values over  15 (Feb
  2010) and  16 (Aug 2010)  observations taken within  approximately 1
  day interval (see Appendix B).  Radio  data are in mJy units. Julian
  dates for radio observations are mean values.}

\begin{tabular}{llccr}
\hline
Filter & Survey & $\lambda_{C}$ & Mag  & Epoch  \\
       &        & [$\mu$m]    & [mag] & [date (JD)] \\
\hline
$Z$   & VVV  & 0.878 & $19.617\pm0.083$ & Mar 30, 2010 (2455285) \\
$Z$   & VVV  & 0.878 & $20.035\pm0.115$ & May 03, 2015 (2457145) \\
$Y$   & VVV  & 1.021 & $18.582\pm0.048$ & Mar 30, 2010 (2455285) \\
$Y$   & VVV  & 1.021 & $18.880\pm0.078$ & May 03, 2015 (2457145) \\
$J$   & VVV  & 1.254 & $17.374\pm0.039$ & Apr 01, 2010 (2455287) \\
$J$   & VVV  & 1.254 & $17.843\pm0.047$ & May 28, 2015 (2457170) \\
$H$   & VVV  & 1.646 & $15.934\pm0.030$ & Apr 01, 2010 (2455287) \\
$H$   & VVV  & 1.646 & $16.557\pm0.047$ & May 28, 2015 (2457170) \\
$K_s$  & VVV & 2.149 & $14.732\pm0.016$ & Apr 01, 2010 (2455287) \\
$K_s$  & VVV & 2.149 & $15.318\pm0.019$ & May 28, 2015 (2457170) \\
W1 & WISE & 3.35 & $13.130\pm0.138$ & Feb 20-21, 2010 (2455249) \\	
W1 & WISE & 3.35 & $13.843\pm0.423$ & Aug 20-22, 2010 (2455430) \\	
W2 & WISE & 4.60 & $12.295\pm0.119$ & Feb 20-21, 2010 (2455249) \\	
W2 & WISE & 4.60 & $13.881\pm0.067$ & Aug 20-22, 2010 (2455430) \\	
W3 & WISE & 11.6 & $\,\,9.481\pm0.206$ & Feb 20-21, 2010 (2455249) \\	
W4 & WISE & 22.1 & $\,\,7.304\pm0.286$ & Feb 20-21, 2010 (2455249) \\ 	 
\hline
Passband &          & Frequency & Flux              &   \\
         &          &          & [mJy]             &   \\
\hline

4.8 GHz & PMN      & 4.8 GHz & $1990\pm99$  & June 1990 (2448057)  \\ 
4.8 GHz & PMN-ATCA & 4.8 GHz & $1041\pm18$  & Nov 9-15, 1992 (2448938) \\
8.6 GHz & PMN-ATCA & 8.6 GHz & $815\pm38$   & Nov 9-15, 1992 (2448938) \\
8.6 GHz & VLBI     & 8.6 GHz & $1463\pm225$ & Dec 12, 2009 (2455177) \\
\hline
\end{tabular}
\end{table*}

\section{Observations and archive data}
\label{sec:obs}

The   VVV   observational   strategy   consists   in   two   sets   of
quasi-simultaneous   $ZY$  and   $JHK_{\rm  s}$   photometry,  and   a
variability  campaign in  the  $K_{\rm s}$-band  with $50-200$  epochs
carried out over  many years ($2010-2016$).  The strategy  of the VVVX
Survey is similar consisting of $JHK_{\rm  s}$ photometry plus 3 to 10
epochs in $K_{\rm s}$-band.

VVV-WIT-04 is located in the VVV {\it tile} d133, towards the Galactic
disk.  In  particular, $ZY$ data for  this tile were collected  on Mar
30, 2010 and May 03, 2015  while $JHK_{\rm s}$ observations were taken
on Apr  01, 2010 and May  28, 2015 (see  Table 1). In addition  to the
colour data, a total of 63 $K_{\rm s}$-band observations spanning from
Mar 31, 2010 to Apr 28, 2018 were also taken with irregular cadence.

\begin{figure*}
\centering
\includegraphics[scale=0.85]{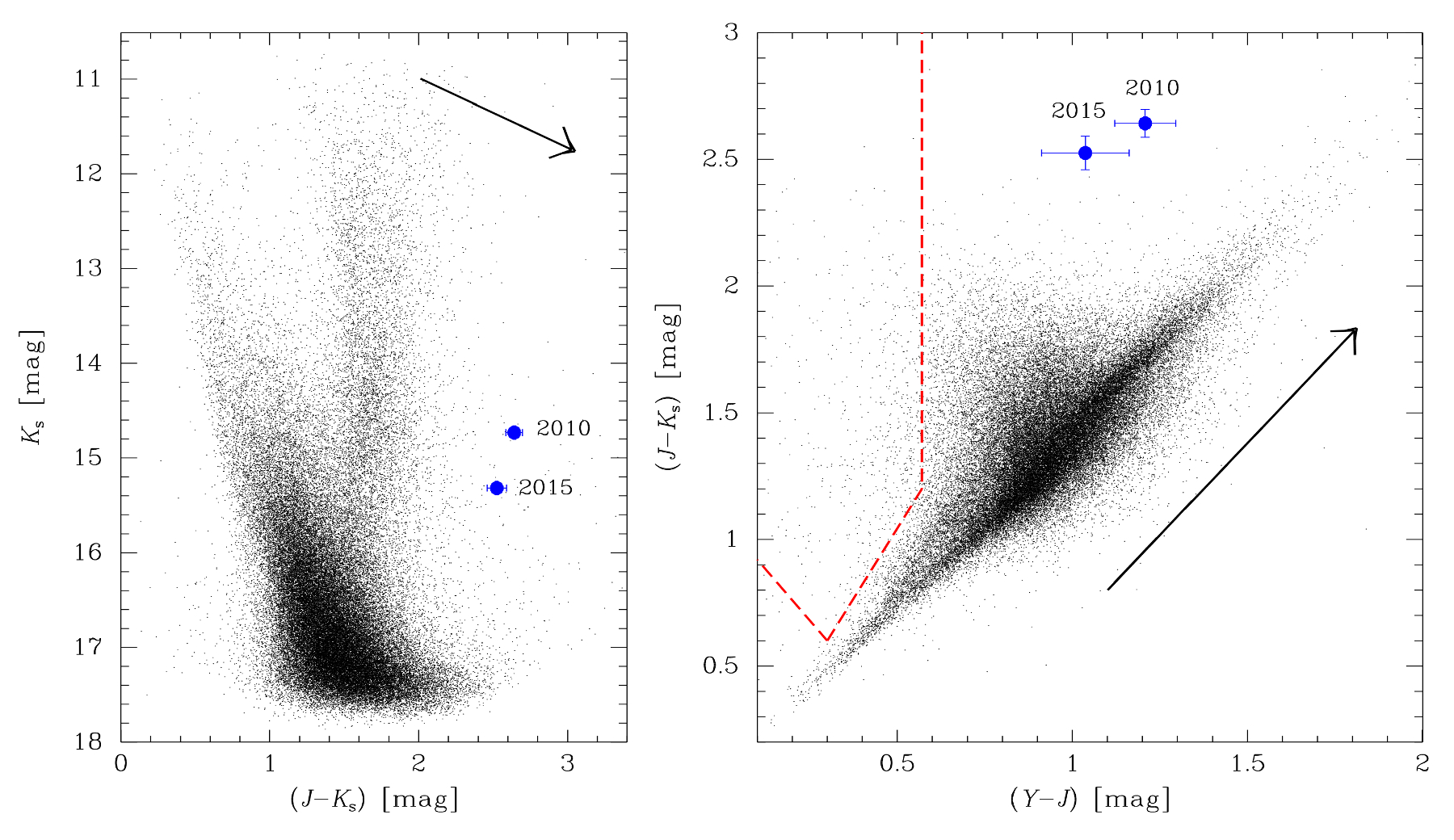}
\caption{$K_{\rm s}$ vs.  $(J-K_{\rm s})$ CMD (left-panel) and $(Z-Y)$
  vs. $(J-K_{\rm s})$ CCD (right-panel)  for stellar sources within 10
  arcmin  of  the  target  position. The  magnitudes  and  colours  of
  VVV-WIT-04 in  year 2010 and 2015  are shown in both  panels as blue
  circles.   The reddening  vector  associated with  an extinction  of
  $A_V=6.52$~mag (see Section 2), based on the relative extinctions of
  the  VISTA  filters,  and assuming  the  \citep{1989ApJ...345..245C}
  extinction law,  is also shown  in both  panels.  In the  CCD dashed
  lines  mark  the  region  populated  by  quasars  found  behind  the
  Magellanic     Clouds     using    VISTA     data     \citep[adapted
    from][]{2016A&A...588A..93I}.   The   colors  of   VVV-WIT-04  are
  consistent with a very reddened quasar.}
\label{fig:cmd}
\end{figure*}

\begin{figure}
\centering
\includegraphics[scale=0.57]{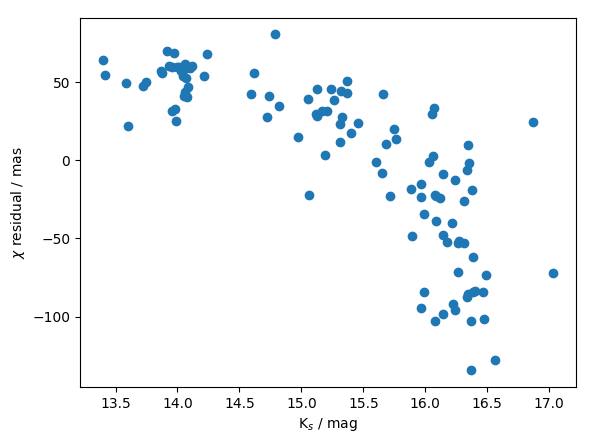}
\caption{Distribution  of the  residuals  to the  proper  motion as  a
  function of the magnitude for the $K_{\rm S}$-band data. It suggests
  contamination  by  a blended  source.   When  VVV-WIT-04 is  in  the
  high-stage  it dominates  the  position while  the contamination  is
  stronger when VVV-WIT-04  is faint, moving the  centroid towards the
  position of the contaminator.}
\label{fig:pm}
\end{figure}

The standard VVV data are based on aperture photometry provided by the
Cambridge  Astronomical Survey  Unit (CASU)  on the  stacked VVV  tile
images \citep[see][for  details]{2012A&A...537A.107S}. Due to  a high
crowding in  the inner  disk --  where VVV-WIT-04  is located  -- both
colour and variability data presented here are based on PSF photometry
performed  on  the  VVV images  \citep[e.g.,  ][]{2017A&A...608A.140C,
  2018A&A...537A.107S}, unlike  the 2010-2013 VVV CASU  data presented
in  \cite{2015ATel.8456....1S}.  The  $K_{\rm s}$-band  light-curve of
VVV-WIT-04  combining  PSF data  from  the  VVV  and VVVX  surveys  is
presented in Fig.~\ref{fig:lcurve}.

VVV-WIT-04  is located  at  coordinates  RA, DEC  (J2000)=15:15:12.69,
$-$55:59:32.78,   corresponding  to   $l,b$=$-$37.869,1.432~deg.   The
position coincides within $\sim 0.2$  arcsec with the radio source PMN
J1515$-$5559                                                  \citep[=
  LQAC\_228-055\_001,][]{1994ApJS...91..111W,2015A&A...583A..75S,
  2018A&A...614A.140G}. Precise coordinates  for J1515$-$5559 from the
Very-Long-Baseline     Interferometry    (VLBI)     Source    Position
Catalogue\footnote{http://astrogeo.org/vlbi/solutions/rfc\_2019a/} are
RA,DEC (J2000)  = 15:15:12.672880,  $-$55:59:32.83821, with  errors in
the  coordinates  as $\sigma_{\rm  RA}$,  $\sigma_{\rm  DEC} =  0.67$,
$0.27$~mas \citep[][and references therein]{2019MNRAS.485...88P}.

A false-color image of the VVV-WIT-04 area produced from the $JHK_{\rm
  s}$ 2010 images is shown in Fig.~\ref{fig:image}: VVV-WIT-04 appears
as a faint point source, much redder than the surrounding field stars.
According to  the VVV extinction maps  \citep{2018A&A...616A..26M} the
region has  a total extinction of  $A_{Ks}=0.77$~mag, corresponding to
$A_V=6.52$~mag, assuming the law of \cite{1989ApJ...345..245C}.  These
values  are  similar  to those  in  \cite{2011ApJ...737..103S},  where
$A_K=0.73$~mag and $A_V=6.65$ mag.

An  archive  search  at  the   VVV-WIT-04  position  resulted  in  few
measurements  at longer  wavelengths.  Two  sets of  observations with
Wide-field Infrared  Survey Explorer  (WISE) were secured  on February
and August  2010 \citep{2012wise.rept....1C,  2013yCat.2328....0C} the
latter  being  simultaneous  with  our VVV  data.   In  the  following
Sections as  well as in  Table 1 the  WISE magnitudes are  mean values
over a dozen observations taken within approximately 1 day interval by
this  satellite  (see   Fig.~\ref{fig:lc_wise}).   The  complete  WISE
dataset is presented  in the Appendix A.  Besides  the measurements in
the VLBI  Source Position Catalogue  \citep{2019MNRAS.485...88P} taken
in Dec  2009 at  8.6~GHz, PMN  J1515$-$5559 was  also observed  by the
Parkes-MIT-NRAO (PMN) Survey \citep{1994ApJS...91..111W} at 4.8 GHz in
June 1990,  and by  the Australia  Telescope PMN  (ATCA-PMN) Follow-up
Survey at 4.8 and 8.6~GHz in Nov 1992 \citep{2012MNRAS.422.1527M}.

\section{Discussion}

VVV-WIT-04  was  found  serendipitously   during  a  search  for  high
amplitude        variables        in        the        VVV        data
\citep{2013A&A...554A.123S,2016ATel.8602....1S}.     As    shown    in
\cite{2015ATel.8456....1S},  the   $K_{\rm  s}$-band   light-curve  of
VVV-WIT-04  covering  2010$-$2013  seasons   is  highly  variable  and
increases in brightness by $\Delta  K_{\rm s}>2.5$ mag during the late
2012  and  the  beginning  of  the 2013  season,  peaking  at  $K_{\rm
  s}=13.4$~mag on May  2013.  After this event,  instead of decreasing
steadily as expected  for a putative outburst (or  even a microlensing
event),  the 2013$-$2018  light-curve shows  an irregular  variability
pattern,  going  fainter than  $K_{\rm  s}=16$~mag  in 2014  and  then
presenting a second peak on May 2015 at $K_{\rm s}=15.2$~mag.  On June
2017 the object  is as faint as $K_{\rm s}=17$~mag,  thus presenting a
total variation of  $\Delta K_{\rm s}>3.6$ mag over the  nine years of
the VVV and VVVX coverage.

\begin{figure}
\centering
\includegraphics[scale=0.8]{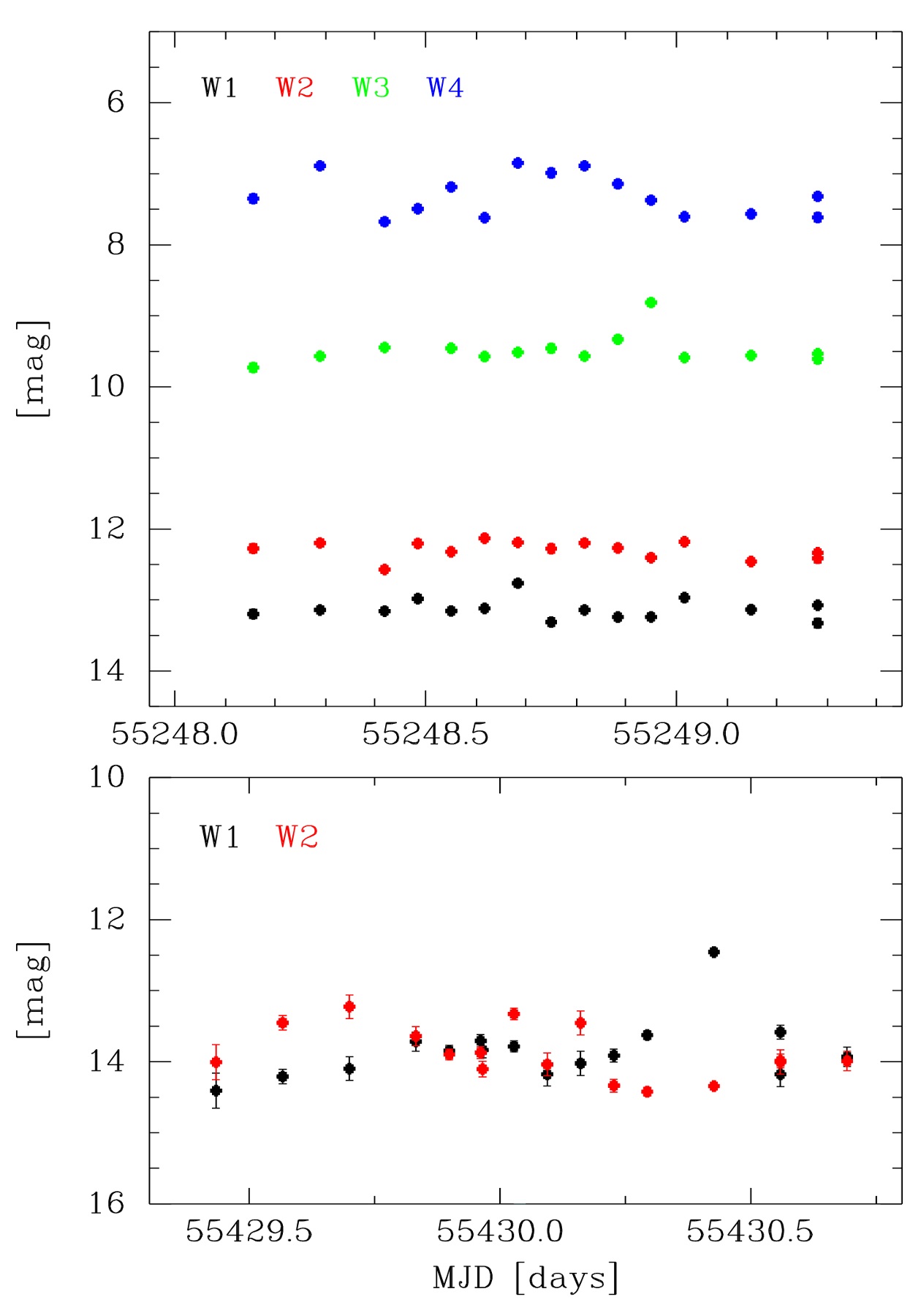}
\caption{WISE light-curves of VVV-WIT-04  within about 1 day coverage.
  Top panel:  Feb 2010 data.  Bottom  panel: Aug 2010 data.   For some
  data-points, especially in the top panel, the error bars are smaller
  than the symbols.}
\label{fig:lc_wise}
\end{figure}

The   colour  of   VVV-WIT-04   also  varies   in   time.   In   2010,
$(Y-J)=1.21$~mag and  $(J-K_{\rm s})=2.64$~mag.   Assuming the  law of
\cite{1989ApJ...345..245C} and  $A_V=6.52$~mag, $(Z-Y)_0=0.49$~mag and
$(J-K_{\rm  s})_0=1.58$~mag.   Later  in  2015,  $(Y-J)=1.03$~mag  and
$(J-K_{\rm  s})=2.52$~mag.    A  $K_{\rm  s}$  vs.    $(J-K_{\rm  s})$
colour-magnitude  diagram  (CMD) and  a  $(Y-J)$  vs. $(J-K_{\rm  s})$
colour-colour diagram (CCD) for stellar sources within 10 arcmin radii
of around  the target position  are shown in  Fig.~\ref{fig:cmd}. Both
diagrams show  that VVV-WIT-04 does  not have typical star  colors. In
fact, the VVV  colors of VVV-WIT-04 are in full  agreement with a very
reddened  quasar, similar  to  the ones  found  behind the  Magellanic
Clouds by  \cite {2016A&A...588A..93I} using VISTA  data, when applied
the extinction  of $A_V=6.52$~mag  towards the position  of VVV-WIT-04
(see  Section 2).  Its WISE  colors are  also consistent  with an  AGN
\cite[e.g., ][]{2012MNRAS.426.3271M,2019A&A...622A..29M}.

Proper   motions  from   the   VVV   Infrared  Astrometric   Catalogue
\citep[VIRAC, ][]{2018MNRAS.474.1826S} show that  the residuals to the
proper motion  varies as a function  of the magnitude for  the $K_{\rm
  s}$-band   observations   by   up   to  $100$~mas,   as   shown   in
Fig.~\ref{fig:pm}.   That  correlation  suggests  contamination  by  a
blended,  faint  source.  When  VVV-WIT-04  is  in the  high-state  it
dominates the target position. On the other hand, when it is faint the
contamination is stronger, moving the centroid towards the position of
the blended contaminant source.

The  WISE observations  also  present variations  in  the mid-IR  (see
Fig.~\ref{fig:lc_wise}).   In Feb  2010  the object  is  seen at  mean
magnitude  $W1=13.10$~mag  with  $(W1 -  W2)=0.84$~mag  compared  with
$W1=13.84$~mag with $(W1 - W2)=-0.04$~mag in Aug 2010.

\section{Possible interpretations}

Quasi-stellar radio sources  - Quasars - are  luminous active galactic
nuclei (AGN). These extragalactic  objects are intrinsically blue, but
due to  local or Galactic  absorption, sometimes appear  as red(dened)
point sources, closely  mimicking a distant star.  Within the Quasar's
zoo are the Optically Violent Variable (OVV) quasars, which are a type
of rare,  highly variable  quasars, proposed to  be unified  under the
class of Flat  Spectrum Radio Quasars (FSRQ).   OVVs are characterized
by very rapid  variability, high and variable polarization  as well as
high  brightness  temperatures   \citep{1995PASP..107..803U}.  A  well
studied case  is the  OVV 3C 279,  with multiwavelength  coverage over
many         years          \citep[][1990$-$2002,         2008$-$2014,
  respectively]{2007AJ....133.2866K,2018MNRAS.479.2037P}.    In    the
optical and near-IR,  3C 279 presents variations as large  as 4 mag on
different timescales.

Close to the OVVs are the BL Lac objects, which are also variable AGNs
presenting a spectral energy distribution  (SED) similar to FSRQs.  BL
Lacs and OVVs are blazar sub-types, which embrace all quasars with the
relativistic  jet  closely  aligned  to  the  line  of  sight  of  the
observer. Compared  with the OVVs,  BL Lac objects are  generally less
luminous  and present  a  relatively featureless  spectrum, with  weak
emission or  absorption lines.   In blazars  both optical  and near-IR
variability  time-scales  depend on  the  distance  from the  emitting
region  to the  central engine  and range  from months  to hours,  the
latter indicating  that the source is  compact \citep[][and references
  therein]{2011MNRAS.411..901G, 2011MNRAS.414.2674G}.

VVV-WIT-04 is a point source located $\sim 0.20$~arcsec apart from the
position   of    PMN   J1515-5559.     Catalogued   as    the   quasar
LQAC\_228-055\_001    \citep{2015A&A...583A..75S,2018A&A...614A.140G},
the  archive  radio  data  of   PMN  J1515-5559  are  consistent  with
non-thermal radiation from  a compact radio source as  expected for an
AGN. As discussed  in Section 3 (see  also Figures~\ref{fig:image} and
\ref{fig:cmd}) the colour of VVV-WIT-04  is not a typical star colour,
but rather it is in agreement  with a very reddened quasar \cite[e.g.,
][]{2012MNRAS.426.3271M,2016A&A...588A..93I,2019A&A...622A..29M},
leading us to  suggest that VVV-WIT-04 has an  extragalactic origin as
the near-IR counterpart of PMN J1515-5559.

In the scenario, the near-IR counterpart is highly variable in time as
shown by  our VVV  light-curve as  well as the  WISE archive  data. In
particular,  the  VVV  light-curve  resembles  the  ones  obtained  by
\citet[][see  their   Fig.~3]{2018MNRAS.479.2037P}  for  the   OVV  3C
279. However, 3C 279 is observed  to vary also at optical wavelengths,
as Optically Violent  Variable Quasar should behave,  while no optical
date are  available to verify  the behaviour of VVV-WIT-04  at shorter
wavelengths. The absence  of optical data is probably due  to the high
extinction. In fact, for a galaxy  behind the MW, the total extinction
as calculated by the VVV maps is probably underestimated.

Our  source has  an  observed magnitude  of $K_s  =  16.5$~mag in  its
"quiescent  phase", which  corresponds  to a  derredened magnitude  of
$K_{\rm s} \sim 15$ mag, assuming an extinction of $A_{Ks} = 1.54$~mag
(with the caveat cited above). By  comparing this magnitude to that of
3C  279  \citep[$K$=10.9 mag  from  the  2MASS point  source  catalog;
][]{2003yCat.2246....0C}  we can  infer that  VVV-WIT-04 is  6.5 times
more   distant.   Based  on   the   WMAP   nine-year  model   cosmolgy
\citep{2012AAS...22050403H},   and   on   a  redshift   of   $z=0.536$
\citep{1996ApJS..104...37M}, the luminosity distance to 3C 279 is 3.13
Gpc.  Were VVV-WIT-04 to have the same luminosity as this prototypical
OVV   QSO,  its   magnitude   would   imply  a   redshift   of  $z   =
2.46$. Therefore, it  is reasonable to assume that  the variability we
detect in  the near infrared  could simply be the  optical variability
shifted towards longer wavelengths due  to the recessional velocity of
the source.

Previous interpretations of VVV-WIT-04 as  a transient event such as a
nova or even a supernova \citep{2015ATel.8456....1S} do not agree with
the current data, especially because of the irregular behaviour during
seasons 2013$-$2018  -- as  example of the  secondary peak  of $\Delta
K_{\rm s}  \sim$1~mag observed on May  2015 -- since the  remnant of a
nova or  a supernova is expected  to decline in brightness  slowly and
steadily with time. That would not  be the first case where a variable
quasar  is misinterpreted  as a  high amplitude  stellar source.   For
instance,  J004457+4123 \citep[=  Sharov 21,  ][]{1998AstL...24..445S}
was first announced as a remarkable nova in M31 and later confirmed as
a     background     quasar     with     a     strong     UV     flare
\citep{2010A&A...512A...1M}.

We have also considered other kinds  of sources highly variable in the
near-IR.   Some  microlensing  events,  for example,  can  have  large
amplitudes.  The amplitude  of a microlensing event is  related to the
impact    parameter    \citep{1986ApJ...304....1P},    therefore,    a
considerable increase in  the brightness of a source  can be explained
with  this effect.   In this  case, the  curve may  resemble a  binary
microlensing  event  due  to  the two  most  pronounced  peaks  around
MJD$\sim$56400  and  MJD$\sim$57200.   To evaluate  this  scenario  we
fitted the light-curve using the python Light-curve Identification and
Microlensing  Analysis  \citep[PyLima, ][]{2017AJ....154..203B}.   The
fit does not follow the observational  data neither in the base (which
is not  constant) nor during  the increases in  brightness.  Moreover,
colour changes  are not expected during  microlensing events, contrary
to  the observed  in VVV-WIT-04.   For these  reasons we  disfavor the
possible explanation of this object as a microlensing event.

\section{Conclusions}

We have presented VVV-WIT-04, a  variable source identified by the VVV
survey towards the  Galactic disk at the position of  the radio source
PMN J1515-5559.  Based on VVV/VVVX variability, multicolor, and proper
motion data our analysis suggests that VVV-WIT-04 has an extragalactic
origin   as  the   near-IR   counterpart  of   the  radio-source   PMN
J1515$-$5559, with  characteristics of  an Optically  Violent Variable
(OVV)  quasar.  The near-IR  variability  can  be interpreted  as  the
redshifted  optical  variability.   Residuals  to  the  proper  motion
suggest that  VVV-WIT-04 is blended  with a nearby source,  probably a
faint star in  the foreground Milky Way  disk.  Alternative scenarios,
including a  transient event such as  a nova or supernova  outburst as
proposed by \cite{2015ATel.8456....1S}, or  even a binary microlensing
event  have also  been discussed  and are  not in  agreement with  the
currently  available  data,  including  the  variability  pattern  and
colours, which disfavors all the listed hypotheses.

The   absence  of   spectroscopic  information   makes  difficult   to
unequivocally classify  VVV-WIT-04 among  the AGN  variable sub-types,
since  the classification  is also  based on  spectral features.   For
instance,  OVVs,  BL  Lacs  or even  UltraLuminous  InfraRed  Galaxies
(ULIRGs)   could  present   similar  variability   behaviour  in   the
optical/near-IR, despite the differences in the luminosity and spectra
\citep[e.g.,
][]{2006asup.book..285L,2011MNRAS.414.2674G,2019MNRAS.483L..17D,2019arXiv190611339G}.

Compact radio sources are distributed over the whole celestial sphere,
including     towards     the      Galactic     plane     \citep[e.g.,
][]{2019MNRAS.485...88P}.  Similar  to VVV-WIT-04 (=  PMN J1515-5559),
other violent variable near-IR counterparts of radio sources should be
present  in the  database  of recent  completed  \citep[e.g., VVV  and
  UKIDSS-GPS,  ][]{2008MNRAS.391..136L} and  ongoing  (e.g., VVVX)  IR
multiepoch surveys  of the inner  Galaxy. A search for  high amplitude
near-IR variability at the position  of radio sources in these surveys
should reveal new interesting objects as is the case of VVV-WIT-04.

\section*{Acknowledgements}

We gratefully acknowledge  the use of data from the  ESO Public Survey
program IDs 179.B-2002 and 198.B-2004  taken with the VISTA telescope,
and data products from the  Cambridge Astronomical Survey Unit (CASU).
This  publication  makes use  of  data  products from  the  Wide-field
Infrared Survey Explorer,  which is a joint project  of the University
of    California,    Los    Angeles,   and    the    Jet    Propulsion
Laboratory/California Institute of Technology,  funded by the National
Aeronautics  and Space  Administration.  R.K.S.   acknowledges support
from     CNPq/Brazil     through    projects     308968/2016-6     and
421687/2016-9.  P.W.L.   is  supported  by  STFC   Consolidated  Grant
ST/R000905/1.   Support  for the  authors  is  provided by  the  BASAL
CONICYT  Center for  Astrophysics and  Associated Technologies  (CATA)
through  grant   AFB-170002,  and   the  Ministry  for   the  Economy,
Development,  and Tourism,  Programa  Iniciativa Cient\'ifica  Milenio
through  grant  IC120009,  awarded  to  the  Millennium  Institute  of
Astrophysics (MAS).  D.M.  acknowledges  support from FONDECYT through
project Regular \#1170121.







\appendix

\section{VVV-WIT-04 $K_{\rm  s}$-band data}

Here we  present the  PSF $K_{\rm  s}$-band data-points  of VVV-WIT-04
available from VVV/VVVX and used to build the light-curve presented in
Fig.  1. There  is a total of 111 data-points  spanning from March 31,
2010 to April  3, 2018. The data-point number of  data-points (111) is
larger than  the observed  epochs (63) because  the PSF  photometry is
performed on the individual VISTA  {\it pawprint} images instead of on
the         final        VISTA         {\it        tile}         image
\citep{2012A&A...537A.107S,2015A&A...575A..25S}.

\hspace{12 pt}

\begin{table}
\centering
\begin{tabular}{cccc}
MJD    & $K_{\rm s}$-band & MJD    & $K_{\rm s}$-band \\
(days) & (mag)          & (days) & (mag)           \\
\hline
\texttt{55287.2971} & \texttt{14.739\,$\pm$\,0.016} & \texttt{56459.0313} & \texttt{13.974\,$\pm$\,0.010}\\
\texttt{55287.2978} & \texttt{14.725\,$\pm$\,0.014} & \texttt{56459.0714} & \texttt{13.980\,$\pm$\,0.013}\\
\texttt{55374.1623} & \texttt{15.884\,$\pm$\,0.041} & \texttt{56459.0866} & \texttt{13.917\,$\pm$\,0.015}\\
\texttt{55374.1627} & \texttt{15.962\,$\pm$\,0.053} & \texttt{56459.0869} & \texttt{13.998\,$\pm$\,0.010}\\
\texttt{55422.0077} & \texttt{16.145\,$\pm$\,0.052} & \texttt{56460.0773} & \texttt{14.089\,$\pm$\,0.012}\\
\texttt{55422.0081} & \texttt{16.063\,$\pm$\,0.054} & \texttt{56460.0777} & \texttt{14.101\,$\pm$\,0.011}\\
\texttt{55423.0662} & \texttt{15.965\,$\pm$\,0.050} & \texttt{56461.1928} & \texttt{13.954\,$\pm$\,0.013}\\
\texttt{55423.0666} & \texttt{15.968\,$\pm$\,0.050} & \texttt{56461.1932} & \texttt{13.934\,$\pm$\,0.011}\\
\texttt{55424.0597} & \texttt{15.890\,$\pm$\,0.045} & \texttt{56462.2526} & \texttt{13.750\,$\pm$\,0.011}\\
\texttt{55424.0601} & \texttt{15.988\,$\pm$\,0.047} & \texttt{56462.2530} & \texttt{13.722\,$\pm$\,0.010}\\
\texttt{55425.0361} & \texttt{16.387\,$\pm$\,0.075} & \texttt{56469.1893} & \texttt{14.031\,$\pm$\,0.012}\\
\texttt{55425.0366} & \texttt{16.078\,$\pm$\,0.056} & \texttt{56469.1897} & \texttt{14.058\,$\pm$\,0.011}\\
\texttt{55430.0372} & \texttt{15.991\,$\pm$\,0.059} & \texttt{56470.1297} & \texttt{14.046\,$\pm$\,0.011}\\
\texttt{55430.0377} & \texttt{16.080\,$\pm$\,0.054} & \texttt{56470.1301} & \texttt{14.051\,$\pm$\,0.010}\\
\texttt{55787.1163} & \texttt{16.468\,$\pm$\,0.098} & \texttt{56471.0304} & \texttt{14.062\,$\pm$\,0.013}\\
\texttt{55803.0393} & \texttt{16.262\,$\pm$\,0.068} & \texttt{56471.0308} & \texttt{14.071\,$\pm$\,0.013}\\
\texttt{55803.0397} & \texttt{16.240\,$\pm$\,0.072} & \texttt{56472.0554} & \texttt{14.073\,$\pm$\,0.016}\\
\texttt{55817.9926} & \texttt{16.372\,$\pm$\,0.071} & \texttt{56472.0730} & \texttt{14.115\,$\pm$\,0.011}\\
\texttt{55817.9929} & \texttt{16.267\,$\pm$\,0.067} & \texttt{56472.0734} & \texttt{14.071\,$\pm$\,0.014}\\
\texttt{55818.9870} & \texttt{16.310\,$\pm$\,0.068} & \texttt{56473.0201} & \texttt{13.865\,$\pm$\,0.012}\\
\texttt{55818.9874} & \texttt{16.315\,$\pm$\,0.086} & \texttt{56473.0205} & \texttt{13.873\,$\pm$\,0.015}\\
\texttt{55819.9880} & \texttt{16.342\,$\pm$\,0.069} & \texttt{56826.0025} & \texttt{16.243\,$\pm$\,0.095}\\
\texttt{55819.9884} & \texttt{16.338\,$\pm$\,0.065} & \texttt{56826.0029} & \texttt{16.377\,$\pm$\,0.089}\\
\texttt{55822.0042} & \texttt{16.368\,$\pm$\,0.067} & \texttt{56826.9796} & \texttt{16.144\,$\pm$\,0.076}\\
\texttt{55822.0047} & \texttt{16.215\,$\pm$\,0.060} & \texttt{56826.9800} & \texttt{16.268\,$\pm$\,0.086}\\
\texttt{56000.3317} & \texttt{16.144\,$\pm$\,0.072} & \texttt{56827.0046} & \texttt{16.342\,$\pm$\,0.084}\\
\texttt{56000.3320} & \texttt{15.963\,$\pm$\,0.062} & \texttt{56827.0050} & \texttt{16.220\,$\pm$\,0.071}\\
\texttt{56055.3022} & \texttt{16.028\,$\pm$\,0.060} & \texttt{56827.0927} & \texttt{16.057\,$\pm$\,0.100}\\
\texttt{56055.3026} & \texttt{16.070\,$\pm$\,0.059} & \texttt{56827.9883} & \texttt{16.083\,$\pm$\,0.091}\\
\texttt{56068.2887} & \texttt{16.116\,$\pm$\,0.056} & \texttt{56827.9887} & \texttt{16.333\,$\pm$\,0.093}\\
\texttt{56068.2892} & \texttt{16.085\,$\pm$\,0.055} & \texttt{56834.0357} & \texttt{16.475\,$\pm$\,0.080}\\
\texttt{56328.3548} & \texttt{15.651\,$\pm$\,0.044} & \texttt{56837.0159} & \texttt{16.356\,$\pm$\,0.078}\\
\texttt{56328.3551} & \texttt{15.717\,$\pm$\,0.049} & \texttt{56837.0163} & \texttt{16.174\,$\pm$\,0.071}\\
\texttt{56365.2609} & \texttt{15.163\,$\pm$\,0.023} & \texttt{56838.9694} & \texttt{16.381\,$\pm$\,0.078}\\
\texttt{56365.2613} & \texttt{15.127\,$\pm$\,0.028} & \texttt{56838.9698} & \texttt{16.400\,$\pm$\,0.077}\\
\texttt{56371.3482} & \texttt{14.820\,$\pm$\,0.018} & \texttt{57112.1651} & \texttt{15.680\,$\pm$\,0.042}\\
\texttt{56371.3486} & \texttt{14.787\,$\pm$\,0.020} & \texttt{57112.1657} & \texttt{15.759\,$\pm$\,0.055}\\
\texttt{56372.2801} & \texttt{15.060\,$\pm$\,0.032} & \texttt{57122.1988} & \texttt{15.456\,$\pm$\,0.044}\\
\texttt{56372.2805} & \texttt{15.123\,$\pm$\,0.031} & \texttt{57122.1992} & \texttt{15.656\,$\pm$\,0.045}\\
\texttt{56373.2215} & \texttt{15.050\,$\pm$\,0.024} & \texttt{57135.1337} & \texttt{15.745\,$\pm$\,0.056}\\
\texttt{56373.2219} & \texttt{14.974\,$\pm$\,0.023} & \texttt{57135.1341} & \texttt{15.605\,$\pm$\,0.041}\\
\texttt{56374.2747} & \texttt{15.117\,$\pm$\,0.048} & \texttt{57170.1658} & \texttt{15.327\,$\pm$\,0.019}\\
\texttt{56374.2751} & \texttt{15.323\,$\pm$\,0.034} & \texttt{57170.1667} & \texttt{15.310\,$\pm$\,0.017}\\
\texttt{56375.2302} & \texttt{15.365\,$\pm$\,0.034} & \texttt{57172.0094} & \texttt{15.403\,$\pm$\,0.047}\\
\texttt{56375.2306} & \texttt{15.370\,$\pm$\,0.033} & \texttt{57172.0098} & \texttt{15.315\,$\pm$\,0.052}\\
\texttt{56377.1707} & \texttt{14.595\,$\pm$\,0.018} & \texttt{57175.0081} & \texttt{15.189\,$\pm$\,0.023}\\
\texttt{56377.1714} & \texttt{14.621\,$\pm$\,0.024} & \texttt{57175.0086} & \texttt{15.206\,$\pm$\,0.023}\\
\texttt{56388.4126} & \texttt{14.214\,$\pm$\,0.013} & \texttt{57184.9903} & \texttt{15.236\,$\pm$\,0.037}\\
\texttt{56388.4130} & \texttt{14.236\,$\pm$\,0.017} & \texttt{57184.9907} & \texttt{15.260\,$\pm$\,0.038}\\
\texttt{56421.3793} & \texttt{13.412\,$\pm$\,0.010} & \texttt{57570.2188} & \texttt{16.528\,$\pm$\,0.096}\\
\texttt{56421.3797} & \texttt{13.401\,$\pm$\,0.011} & \texttt{57570.2192} & \texttt{16.565\,$\pm$\,0.081}\\
\texttt{56422.3114} & \texttt{13.600\,$\pm$\,0.011} & \texttt{57922.1606} & \texttt{16.451\,$\pm$\,0.071}\\
\texttt{56422.3118} & \texttt{13.585\,$\pm$\,0.015} & \texttt{57924.1200} & \texttt{17.028\,$\pm$\,0.168}\\
\texttt{56458.1027} & \texttt{13.952\,$\pm$\,0.011} & \texttt{57924.1204} & \texttt{16.871\,$\pm$\,0.126}\\
\texttt{56458.1031} & \texttt{13.988\,$\pm$\,0.009} & \texttt{58212.3241} & \texttt{16.486\,$\pm$\,0.106}\\
\texttt{56459.0309} & \texttt{14.009\,$\pm$\,0.011} &                     &                              \\
\end{tabular}
\end{table}

\newpage

\section{WISE data}

\begin{table}
\centering
\begin{tabular}{ccccc}
MJD    & W1    & W2    & W3    & W4    \\
(days) & (mag) & (mag) & (mag) & (mag) \\
\hline
\texttt{55248.1549} & \texttt{13.198\,$\pm$\,0.066} & \texttt{12.275\,$\pm$\,0.055} & \texttt{9.729\,$\pm$\,0.121} & \texttt{7.350\,$\pm$\,0.418}\\  
\texttt{55248.2873} & \texttt{13.141\,$\pm$\,0.055} & \texttt{12.197\,$\pm$\,0.054} & \texttt{9.566\,$\pm$\,0.112} & \texttt{6.889\,$\pm$\,0.220}\\  
\texttt{55248.4196} & \texttt{13.156\,$\pm$\,0.047} & \texttt{12.572\,$\pm$\,0.091} & \texttt{9.444\,$\pm$\,0.121} & \texttt{7.674\,$\pm$\,0.448}\\  
\texttt{55248.4857} & \texttt{12.983\,$\pm$\,0.061} & \texttt{12.204\,$\pm$\,0.066} &              ---             & \texttt{7.493\,$\pm$\,0.343}\\  
\texttt{55248.5519} & \texttt{13.154\,$\pm$\,0.054} & \texttt{12.321\,$\pm$\,0.079} & \texttt{9.456\,$\pm$\,0.106} & \texttt{7.187\,$\pm$\,0.348}\\  
\texttt{55248.6181} & \texttt{13.119\,$\pm$\,0.041} & \texttt{12.130\,$\pm$\,0.035} & \texttt{9.573\,$\pm$\,0.133} & \texttt{7.619\,$\pm$\,0.428}\\  
\texttt{55248.6842} & \texttt{12.764\,$\pm$\,0.045} & \texttt{12.189\,$\pm$\,0.053} & \texttt{9.513\,$\pm$\,0.114} & \texttt{6.848\,$\pm$\,0.228}\\  
\texttt{55248.7505} & \texttt{13.311\,$\pm$\,0.065} & \texttt{12.277\,$\pm$\,0.057} & \texttt{9.457\,$\pm$\,0.095} & \texttt{6.988\,$\pm$\,0.240}\\  
\texttt{55248.8165} & \texttt{13.141\,$\pm$\,0.055} & \texttt{12.197\,$\pm$\,0.054} & \texttt{9.566\,$\pm$\,0.112} & \texttt{6.889\,$\pm$\,0.220}\\ 
\texttt{55248.8828} & \texttt{13.241\,$\pm$\,0.056} & \texttt{12.268\,$\pm$\,0.063} & \texttt{9.330\,$\pm$\,0.112} & \texttt{7.143\,$\pm$\,0.339}\\  
\texttt{55248.9490} & \texttt{13.239\,$\pm$\,0.050} & \texttt{12.403\,$\pm$\,0.051} & \texttt{8.813\,$\pm$\,0.395} & \texttt{7.372\,$\pm$\,0.469}\\  
\texttt{55249.0151} & \texttt{12.968\,$\pm$\,0.046} & \texttt{12.180\,$\pm$\,0.043} & \texttt{9.587\,$\pm$\,0.118} & \texttt{7.605\,$\pm$\,0.480}\\  
\texttt{55249.1474} & \texttt{13.136\,$\pm$\,0.041} & \texttt{12.459\,$\pm$\,0.059} & \texttt{9.558\,$\pm$\,0.128} & \texttt{7.566\,$\pm$\,0.491}\\  
\texttt{55249.2797} & \texttt{13.326\,$\pm$\,0.067} & \texttt{12.411\,$\pm$\,0.077} & \texttt{9.605\,$\pm$\,0.112} & \texttt{7.614\,$\pm$\,0.413}\\  
\texttt{55249.2798} & \texttt{13.074\,$\pm$\,0.038} & \texttt{12.337\,$\pm$\,0.063} & \texttt{9.533\,$\pm$\,0.118} & \texttt{7.318\,$\pm$\,0.386}\\  
\texttt{55429.4337} & \texttt{14.407\,$\pm$\,0.245} & \texttt{14.006\,$\pm$\,0.292} &        ---        &      ---           \\  
\texttt{55429.5660} & \texttt{14.208\,$\pm$\,0.103} & \texttt{13.452\,$\pm$\,0.116} &        ---        &      ---           \\  
\texttt{55429.6984} & \texttt{14.098\,$\pm$\,0.166} & \texttt{13.225\,$\pm$\,0.095} &        ---        &      ---           \\  
\texttt{55429.8307} & \texttt{13.716\,$\pm$\,0.134} & \texttt{13.639\,$\pm$\,0.186} &        ---        &      ---           \\  
\texttt{55429.8967} & \texttt{13.848\,$\pm$\,0.077} & \texttt{13.898\,$\pm$\,0.244} &        ---        &      ---           \\  
\texttt{55429.9628} & \texttt{13.705\,$\pm$\,0.085} & \texttt{13.872\,$\pm$\,0.308} &        ---        &      ---           \\  
\texttt{55429.9630} & \texttt{13.836\,$\pm$\,0.108} & \texttt{14.104\,$\pm$\,0.212} &        ---        &      ---           \\  
\texttt{55430.0290} & \texttt{13.785\,$\pm$\,0.079} & \texttt{13.328\,$\pm$\,0.136} &        ---        &      ---           \\  
\texttt{55430.0951} & \texttt{14.178\,$\pm$\,0.160} & \texttt{14.037\,$\pm$\,0.253} &        ---        &      ---           \\  
\texttt{55430.1614} & \texttt{14.023\,$\pm$\,0.170} & \texttt{13.454\,$\pm$\,0.121} &        ---        &      ---           \\  
\texttt{55430.2274} & \texttt{13.912\,$\pm$\,0.090} & \texttt{14.336\,$\pm$\,0.266} &        ---        &      ---           \\  
\texttt{55430.2937} & \texttt{13.626\,$\pm$\,0.061} & \texttt{14.422 ~~~\,\,\,\,\, } &       ---         &     ---            \\  
\texttt{55430.4260} & \texttt{12.456 ~~~~~        } & \texttt{14.342\,$\pm$\,0.337} &        ---        &      ---           \\  
\texttt{55430.5581} & \texttt{13.583\,$\pm$\,0.098} & \texttt{13.987\,$\pm$\,0.290} &        ---        &      ---           \\  
\texttt{55430.5583} & \texttt{14.177\,$\pm$\,0.172} & \texttt{14.003\,$\pm$\,0.239} &        ---        &      ---           \\  
\texttt{55430.6904} & \texttt{13.931\,$\pm$\,0.136} & \texttt{13.990 ~~~\,\,\,\,\, } &       ---         &     ---            \\  

\end{tabular}
\end{table}

\bsp	
\label{lastpage}
\end{document}